\documentclass[aps,pra,twocolumn,superscriptaddress,showpacs,groupedaddress,noeprint,10pt,footinbib]{revtex4-1}

\usepackage[utf8]{inputenc}
\usepackage[T1]{fontenc}
\usepackage{mathptmx}

\usepackage{amsmath}
\usepackage{amssymb}
\usepackage{graphicx}

\usepackage[export]{adjustbox}
\usepackage{float}
\usepackage{color}
\usepackage{bm}
\usepackage{xcolor}

\usepackage{url}
\usepackage{hyperref}
\hypersetup{
	colorlinks=true,
	urlcolor= blue,
	citecolor=blue,
	linkcolor= blue,
	bookmarks=true,
	bookmarksopen=false,
}

\begin{document}

\title{Terahertz response of gadolinium gallium garnet (GGG) and gadolinium scandium gallium garnet (SGGG)}

\author{Mohsen Sabbaghi}
\email[]{sabbagh2@uwm.edu}
\affiliation{University of Wisconsin-Milwaukee, Milwaukee, Wisconsin, 53211, USA}

\author{George W. Hanson}
\email[]{george@uwm.edu}
\affiliation{University of Wisconsin-Milwaukee, Milwaukee, Wisconsin, 53211, USA}

\author{Michael Weinert}
\email[]{weinert@uwm.edu}
\affiliation{University of Wisconsin-Milwaukee, Milwaukee, Wisconsin, 53211, USA}

\author{Fan Shi}
\email[]{fan.shi@mail.wvu.edu}
\affiliation{West Virginia University, Morgantown, West Virginia, 26506, USA}
\affiliation{Key Laboratory of Computer Vision and System of Ministry of Education, Tianjin University of Technology, Tianjin, 300384, China}

\author{Cheng Cen}
\email[]{cheng.cen@wvu.edu}
\affiliation{West Virginia University, Morgantown, West Virginia, 26506, USA}

\date{\today}

\begin{abstract}
We report the magneto-optical response of Gadolinium Gallium Garnet (GGG) and Gadolinium Scandium Gallium Garnet (SGGG) at frequencies ranging from $300 \, \mathrm{GHz}$ to $1 \, \mathrm{THz}$, and determine the  material response tensor. Within this frequency window, the materials exhibit nondispersive and low-loss optical responses. At low temperatures, significant THz Faraday rotations are found in the (S)GGG samples. Such strong gyroelectric response is likely associated with the high-spin paramagnetic state of the Gd$^{3+}$ ions. A model of the material response tensor is determined, together with the Verdet and magneto-optic constants.
\end{abstract}

\maketitle

\section{Introduction}\label{SEC:INTRODUCTION}

GGG (Gd$_{5}$Ga$_{3}$O$_{12}$) and substituted GGG (e.g., Gd$_{3}$Sc$_{2}$Ga$_{3}$O$_{12}$) belong to the garnet material family, which are described by the general chemical formula A$_{3}$B$_{2}$C$_{3}$O$_{12}$, with A, B and C being metal ions which are trapped inside oxygen dodecahedrals, octahedrals and tetrahedrals, respectively \cite{GELLER, Young_Jin}.

The nominal electronic configuration of the constituent ions of (S)GGG is: Gd$^{3+}$ ([Xe]4$f^{7}$5$d^{0}$6$s^{0}$, O$^{2-}$ ([He]2$s^{2}$2$p^{6}$), Sc$^{3+}$ ([Ar]3$d^{0}$4$s^{0}$) and Ga$^{3+}$ ([Ar]3$d^{10}$4$s^{0}$4$p^{0}$). Among them, only the Gd$^{3+}$ ions have a non-zero magnetic moment \cite{Young_Jin}. Under low magnetic fields, (S)GGG is paramagnetic. When the external field exceeds $1\,\mathrm{T}$, however, a field-induced antiferromagnetic phase can be produced in GGG at temperatures below $1\,\mathrm{K}$ \cite{HOV1980455,Dai_1988,PhysRevLett.73.2500,PhysRevLett.74.2379,PhysRevLett.80.4570,10.1063/1.370392,PETRENKO199941,Petrenko2002,Petrenko_2009,PhysRevB.91.014419,PhysRevD.91.102004,PhysRevLett.114.227203,Paddison179}.

Owing to the closely matched lattice structures, crystalline (S)GGG is widely used as the growth substrate for a general class of iron garnets described by X$_{3}$Fe$_{5}$O$_{12}$ (XIG)  \cite{Dijkkamp_1987,Scheel_2007} with spintronic applications, such as TIG \cite{Tang_2016,Avci_2016,Tang_2017,Quindeau_2017,Wu_2018}, YIG \cite{Bedyukh_1999,Chiang_2002,Pashkevich_2012,Jungfleisch_2013,Tao_2013,Wang_2014,Cornelissen_2015,Sokolov_2016,Stupakiewicz_2016,Holanda_2018,Maehrleineaar5164,Seifert_2018,Liu_2018}, BIG \cite{Kahl_2003,POPOVA_2013,Pohl_2013}, HIG \cite{Kalashnikova_2012}, TbIG \cite{Kumar_2008} and GdIG \cite{Stephan_2016} (X = {Tm}, {Yb}, {Bi}, {Ho}, {Tb} and {Gd}, respectively). The epitaxial strain induced by the (S)GGG substrate, tunable by the B-site substitution, can be used to effectively manipulate the magnetization and magnetic easy-axis of the XIG films. Recently, strain induced out-of-plane ferrimagnetic ordering in TIG films grown on (S)GGG substrates has been utilized to generate room-temperature magnetic proximity effects in topological insulators \cite{Tang_2017}. Similar XIG/GGG heterostructures have also been used to realize spin pumping in single-layer graphene \cite{PhysRevB.95.024408,PhysRevLett.120.097702}.

Garnets, with the highest magneto-optical Verdet constants in bulk media, are the most common materials for Faraday rotators and optical isolators at visible wavelengths \cite{Stadler_2014,Dulal_2016,stadler_hutchings_2018}. In (S)GGG, the paramagnetic Gd$^{3+}$ ions with a large spin (7 unpaired 4f electrons) are effective enablers of strong magneto-optical effects. As a substrate material, (S)GGG based heterostructures may find even more novel photonic applications. For example, alternating deposition of ultra-thin XIG and GGG films has led to the realization of all-garnet magneto-optical photonic crystals (MOPCs) \cite{Lyubchanskii_2003,Lyubchanskii_2004_1,Lyubchanskii_2004_2,Kahl_2004,Rong_2005,Vasiliev_2008,Alam_2009,Wu_2010,Lyubchanskii_2010,Grishin_2012,Alameh_2012,Alam_2015,Chekhov_2015,Rachid_2017,Alam_2018}. Additionally, heterostructures combining (S)GGG-based garnet substrates and 2D quantum materials that exhibit giant Faraday rotations at THz frequencies \cite{Tang_2017,PhysRevB.95.024408,PhysRevLett.120.097702} can potentially lead to broadband magneto-optical devices that cover the whole THz-to-visible frequency range.

While the optical properties of (S)GGG at visible wavelengths have been well studied, their optical and magneto-optical optical responses at THz frequencies are little explored \cite{PODRAZA}. To support the future design of broadband devices based on 2D material/(S)GGG heterostructures, the objective of this work is to measure and model the permittivity and permeability of crystalline (S)GGG for $0.3$--$1\,\mathrm{THz}$ and within the temperature range of $5$--$295\,\mathrm{K}$ where the phonon \cite{PODRAZA} or magnon \cite{PhysRevLett.114.227203} excitations are absent.

As substrate materials used for thin film epitaxy, (S)GGG single crystals often need to undergo thermal treatment in oxygen environment to form an atomically flat surface with uniform surface termination. Since spin properties in correlated oxides are sensitive to small lattice distortions and defect formations that can occur during the thermal annealing process, experiments are performed on both as-grown (untreated) and annealed (S)GGG samples to explore their potential impacts on the magneto-optical responses.

In what follows, we first discuss the gyrotropic response tensors of (S)GGG, and relate the Faraday rotation to material properties. Terahertz time-domain spectroscopy is then discussed, and the measured Faraday rotation presented. The refractive index and gyrotropic elements of the material response tensor is then obtained, as are the Verdet and magneto-optics constants.
\section{Material Response Formalism}\label{SEC:FORMALISM}

In this section, we present the gyroelectric material response tensor, and relate its elements to measured values.

According to the Onsager-Casimir symmetry relations \cite{Onsager_1931,Casimir_1945}, the permittivity and permeability tensors of a crystal should be symmetrical, i.e., for $n,m=x,y,z$
\begin{eqnarray}
\varepsilon_{m,n}\!\left(\omega , \mathbf{k} , \mathbf{B} \right) &=& \varepsilon_{n,m}\!\left(\omega , -\mathbf{k} , - \mathbf{B} \right) \label{ONSAGER_CASIMIR_PERMITTIVITY}
\\[1.0ex]
\mu_{m,n}\!\left(\omega , \mathbf{k} , \mathbf{B} \right) &=& \mu_{n,m}\!\left(\omega , -\mathbf{k} , - \mathbf{B} \right)  \label{ONSAGER_CASIMIR_PERMEABILITY} , 
\end{eqnarray}
where $\varepsilon_{m,n}$ and $\mu_{m,n}$ denote the generic components of the permittivity and permeability tensors, respectively. The variables $\omega$, $\mathbf{k}$ and $\mathbf{B}$ denote, respectively, the angular frequency, wavevector and DC magnetic flux density vector. As a consequence, in the general case of triclinic crystal symmetry, there are 18 complex tensor components to be determined.

We consider a material slab of finite thickness under a $z$-directed magnetic bias whose electromagnetic (EM) response is described by permittivity and permeability tensors (denoted by $\pmb{\varepsilon}$ and $\pmb{\mu}$, respectively) of gyrotropic form, with their components being represented by the matrices
\begin{equation}\label{GYROTROPIC_PERMITTIVITY}
\pmb{\varepsilon} \equiv \begin{bmatrix} \varepsilon_{d} & i \varepsilon_{g} & 0 \\[0.5ex] -i \varepsilon_{g} & \varepsilon_{d} & 0 \\[0.5ex] 0 & 0 & \varepsilon_{a} \end{bmatrix} , \quad \pmb{\mu} \equiv \begin{bmatrix} \mu_{d} & i \mu_{g} & 0 \\[0.5ex] -i \mu_{g} & \mu_{d} & 0 \\[0.5ex] 0 & 0 & \mu_{a} \end{bmatrix},
\end{equation}
where the coordinate system is chosen so that the $z$-axis is perpendicular to the slab. The subscripts $d$, $g$ and $a$ specify the diagonal, off-diagonal, and axial components, respectively. The cubic symmetry of the (S)GGG crystal requires the permittivity and permeability tensors of unbiased (S)GGG to be isotropic. Since the bias is perpendicular to slab, it does not break the isotropy in the $x$-$y$ plane, and thus the in-plane diagonal components of the each of these tensors are expected to remain equal in the presence of bias. However, in general, the axial ($zz$) component of each of these tensors in the presence of the $z$-directed bias will differ from the in-plane diagonal components, i.e., $\varepsilon_{xx} = \varepsilon_{yy} = \varepsilon_{d} \neq \varepsilon_{zz} = \varepsilon_{a}$ and $\mu_{xx} = \mu_{yy} = \mu_{d} \neq \mu_{zz}=\mu_{a}$.

A linearly-polarized normally-incident plane wave can be decomposed into left-handed (LHCP) and right-handed circularly-polarized (RHCP) components of equal amplitude. These components propagate through the gyrotropic substrate according to their corresponding refractive indices (or eigenvalues),
\begin{equation}\label{LHCP_RHCP_REFRACTIVE_INDICES}
n_{\text{\tiny{R}} / \text{\tiny{L}}} \equiv \sqrt{ \left( \varepsilon_{d} \mu_{d} + \varepsilon_{g} \mu_{g}\right) \pm \left( \varepsilon_{d} \mu_{g} + \varepsilon_{g} \mu_{d}\right)}.
\end{equation}
As a result, the LHCP and RHCP components have different phase velocities, and the resulting phase difference causes the polarization of the plane wave to be rotated; a phenomenon known as Faraday rotation (FR). Moreover, in a lossy gyrotropic medium the LHCP and RHCP components are attenuated at different rates upon propagation, such that the polarization state changes from linear to elliptic upon transmission. The resulting degree of ellipticity is referred to as Faraday ellipticity (FE). Excluding interference effects (which can be time-gated out), the evolution of polarization due to transmission through a gyrotropic slab of thickness $d$ is described by \cite{Shimano}
\begin{equation}\label{FARADAY_ROTATION_1}
\frac{\overline{E}_{y}\!\left(\lambda\right)}{\overline{E}_{x}\!\left(\lambda\right)} =  \frac{\sin{ \theta_{\text{\tiny{F}}}} + i \eta_{\text{\tiny{F}}} \cos{ \theta_{\text{\tiny{F}}}}}{\cos{ \theta_{\text{\tiny{F}}}} - i \eta_{\text{\tiny{F}}} \sin{ \theta_{\text{\tiny{F}}}}} = \tan{\!\left[\frac{\pi d}{\lambda} \left[ n_{\text{\tiny{R}}} - n_{\text{\tiny{L}}} \right]\right]}.
\end{equation}
with $\lambda$, $\eta_{\text{\tiny{F}}}$ and $\theta_{\text{\tiny{F}}}$ denoting, respectively, the vacuum wavelength, the Faraday ellipticity, and the angle between the polarization of the linearly-polarized incident EM field and the major axis of the ellipse traced out by the tip of the electric field of the transmitted EM field (the FR).

The gyrotropic response described by Eq.~(\ref{GYROTROPIC_PERMITTIVITY}) can be induced by an external bias, e.g., a perpendicularly-applied static magnetic flux intensity $B_{z}$. In the limit of weak gyrotropy, where the bias is small, i.e., $\left|\varepsilon_{g}\right| \ll \left|\varepsilon_{d}\right|$ and $\left|\mu_{g}\right| \ll \left|\mu_{d}\right|$, the LHCP and RHCP refractive indices differ by a small amount, and the FR is not expected to be large. As a result, Eq.~(\ref{FARADAY_ROTATION_1}) can be simplified to
\begin{equation}\label{FARADAY_ROTATION_2}
\frac{\overline{E}_{y}\!\left(\lambda\right)}{\overline{E}_{x}\!\left(\lambda\right)} \cong \theta_{\text{\tiny{F}}} + i \eta_{\text{\tiny{F}}} \cong \frac{\pi d}{\lambda} \left[ n_{\text{\tiny{R}}} - n_{\text{\tiny{L}}} \right],
\end{equation}
and the combination of Eqs.~(\ref{LHCP_RHCP_REFRACTIVE_INDICES}) and (\ref{FARADAY_ROTATION_2}) results in
\begin{equation}\label{FARADAY_ROTATION_IN_THE_LIMIT_OF_WEAK_GYROTROPY}
\theta_{\text{\tiny{F}}} + i \eta_{\text{\tiny{F}}} \cong \frac{\pi d}{\lambda} \left[ \frac{\varepsilon_{g} \mu_{d} + \varepsilon_{d} \mu_{g}}{\sqrt{\varepsilon_{d}\mu_{d}}} \right].
\end{equation}
Assuming the medium to exhibit gyroelectric response \footnote{The assumption of gyroelectric ($\varepsilon_{g} \neq 0$ and $\mu_{g}=0$) response can be justified by noting the experimental observation that the Gd moments can order antiferromagnetically at low temperatures ($T< 0.2\, \mathrm{K}$) in weak magnetic fields \cite{Paddison179}. The magnetic properties imply that the Gd moments are weakly coupled to each other and to the lattice, and thus that $\mu_{g} \sim 0$. Our DFT calculations for GGG support these conclusions: The weak interactions result from both the relatively large Gd-Gd separations and, more importantly, from the fact that the Gd 4$f$ states are effectively core-like, with little hybridization with the GGG valence states, which again argues for $\mu_{g}\sim 0$. The local exchange fields due to the Gd 4$f^{7}$ moments cause a polarization of the Gd valence and conduction states, but the quantization direction is determined by the external field.}, the magnetic bias results in a non-diagonal permittivity tensor, while the permeability tensor remains diagonal, i.e., $\varepsilon_{\text{g}} \neq 0$ and $\mu_{g} = 0$. In this case, Eq.~(\ref{FARADAY_ROTATION_IN_THE_LIMIT_OF_WEAK_GYROTROPY}) reduces to
\begin{equation}\label{FARADAY_ROTATION_IN_THE_LIMIT_OF_WEAK_GYROELECTRICITY_1}
\theta_{\text{\tiny{F}}} + i \eta_{\text{\tiny{F}}} \cong \frac{\pi d}{\lambda} \varepsilon_{g} \sqrt{\frac{ \mu_{d}}{\varepsilon_{d}}}.
\end{equation}
Therefore, the off-diagonal elements of the permittivity tensor of a gyroelectric medium can be studied through FR and FE measurements. Throughout this work, the weak gyrotropy limit is assumed, in which only the terms linear in bias are retained. Since the bias-dependence of $\varepsilon_{d}$ in Eq.~(\ref{FARADAY_ROTATION_IN_THE_LIMIT_OF_WEAK_GYROELECTRICITY_1}) manifests itself through terms quadratic in bias and higher, it can be replaced with the permittivity of unbiased substrate, $\varepsilon_{d,0}$,
\begin{equation}\label{PERMITTIVITY_IN_TERMS_OF_THE_REAL_AND_IMAGINARY_PARTS_OF_REFRACTIVE_INDEX}
\varepsilon_{d} \cong \varepsilon_{d,0} = \left(n - i \kappa \right)^{2},
\end{equation}
with $n$ and $\kappa > 0$ being the refractive index and extinction constant of the unbiased substrate. Therefore, Eq.~(\ref{FARADAY_ROTATION_IN_THE_LIMIT_OF_WEAK_GYROELECTRICITY_1}) reduces to
\begin{equation}\label{FARADAY_ROTATION_IN_THE_LIMIT_OF_WEAK_GYROELECTRICITY_2}
\theta_{\text{\tiny{F}}} + i \eta_{\text{\tiny{F}}} \cong \frac{\pi d}{\lambda} \frac{\sqrt{1 + \chi}}{n - i \kappa} \varepsilon_{g},
\end{equation}
where the relative permeability $\mu_{d}$ is expressed in terms of the magnetic susceptibility of the substrate, assumed to be real, and defined as $\chi \equiv \mu_{d} - 1$. At frequencies close to electron paramagnetic resonance (EPR), the magnetic susceptibility of (S)GGG, in its paramagnetic phase, has an imaginary part, which manifests itself through microwave loss \cite{Dutoit_1974,Adam_1976,Bedyukh_1999}. The EPR resonance frequency is proportional to the DC magnetic bias, and for a magnetic bias as high as $0.4\,\mathrm{T}$, the EPR peak occurs at around $100 \, \mathrm{GHz}$. In this work, we focus on the low frequency range between 0.3-1 THz, in which the samples measured exhibit almost negligible loss. The higher frequency properties, involving strong coupling to the phonon modes in (S)GGG, will be discussed elsewhere. Within this frequency range, the imaginary part of susceptibility can be approximated to be proportional to the bias, and its contribution to the left-hand side of Eq. (9) is through terms which are quadratic in bias or higher. As a result, within the $0.3$--$1\, \mathrm{THz}$ frequency range, and for magnetic biases up to $0.4\,\mathrm{T}$, the magnetic susceptibility of (S)GGG can be approximated by its purely-real DC value (i.e., we can neglect dispersion), and the real and imaginary parts of the off-diagonal component of the permittivity tensor can be obtained from experimental data as
\begin{eqnarray}
\mathrm{Re}{\left[ \varepsilon_{g} \right]} &\cong& \frac{\lambda}{\pi d} \left[ \frac{n \theta_{\text{\tiny{F}}} + \kappa \eta_{\text{\tiny{F}}}}{ \sqrt{1 + \chi}} \right], \label{REAL_PART_OF_THE_OFFDIAGONAL_COMPONENT_OF_THE_PERMITTIVITY_TENSOR}
\\[1.0ex]
\mathrm{Im}{\left[ \varepsilon_{g} \right]} &\cong& \frac{\lambda}{\pi d} \left[ \frac{n \eta_{\text{\tiny{F}}} - \kappa \theta_{\text{\tiny{F}}}}{\sqrt{1 + \chi}} \right]\label{IMAGINARY_PART_OF_THE_OFFDIAGONAL_COMPONENT_OF_THE_PERMITTIVITY_TENSOR}.
\end{eqnarray}
Therefore, in order to determine the permittivity tensor elements, it is necessary to measure the refractive index, $n$, extinction coefficient, $\kappa$, Faraday rotation, $ \theta_{\text{\tiny{F}}}$, Faraday ellipticity, $ \eta_{\text{\tiny{F}}}$, and magnetic susceptibility, $\chi$. We do not obtain $\varepsilon_{a}$ or $\mu_{a}$ in this work.

\section{Transmission measurements of (S)GGG substrates} \label{SECTION_II}

\subsection{Terahertz time-domain spectroscopy}\label{SUBSEC_II_I}

\begin{figure}[ht!]
	\begin{center}
		\includegraphics[clip,width=0.95\columnwidth]{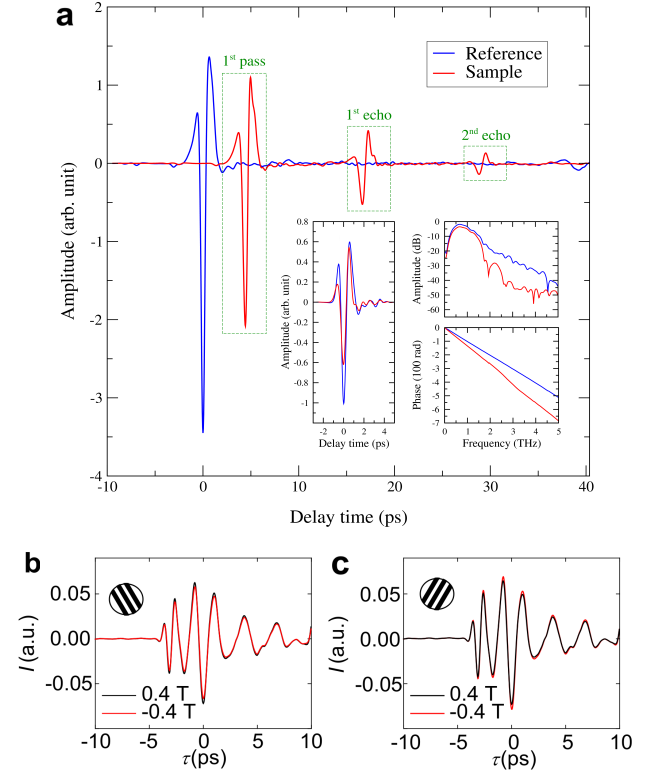}\\
		\caption{(a) The electric field amplitude measured in the time domain in the presence (absence) of the substrate is labeled as``sample'' (``reference''). The echo pulse seen in the reference spectrum at  around $t = 38 \, \mathrm{ps}$ is due to the presence of the thicker quartz windows of the cryostation \cite{Withawat_2008}. The insets in panel (a) show the windowed ``first-pass'' and the ``reference'' pulses in the time domain, and the amplitude and phase of their discrete Fourier transform. Panels (b) and (c) show the detected signal for the two wire-grid polarizer angles of $\pm 30^{\circ}$ for the biased sample. The Faraday rotation is determined through detecting small changes to these amplitudes induced by magnetic bias.} 
		\label{FIG_THZ_TDS}
	\end{center}
\end{figure}

Samples used in this experiment are $\langle 111 \rangle$-oriented single crystal GGG and SGGG with a nominal thickness of $d =\! 0.5 \, \mathrm{mm}$. Variable-temperature THz transmission and Faraday rotation measurements are performed in a cryostat using THz time-domain spectroscopy (THz-TDS). Figure~\ref{FIG_THZ_TDS} shows the typical transient THz waveform transmitted through the sample. The transmission spectrum is obtained by comparing the Fourier transforms of the time-domain signals measured with and without the sample. The time-domain sample signal is truncated to remove the interference effects associated with the echo pulses \cite{Van_Exter_1989}. Faraday rotation is characterized by comparing the transmitted signals measured with two different detector polarizer angles ($\pm30$ degrees).

\subsection{Transmission measurement of unbiased substrates: refractive index and absorption}\label{SUBSEC_II_II}

For weak magnetic bias, the diagonal elements of the permittivity tensor can be obtained from the complex refractive index, $\tilde{n} = n - i \kappa$, of the unbiased sample. The transmission-mode THz-TDS measurements make it possible to extract $\tilde{n}$ from the following equation \cite{Withayachumnankul}
\begin{equation}\label{TRANSMISSION_MODE_THZ_TDS}
\frac{\overline{E}_{\mathrm{sam}}\left(\lambda\right)}{\overline{E}_{\mathrm{ref}}\left(\lambda\right)} \cong \frac{4 \tilde{n}}{ \left( 1 + \tilde{n} \right)^{2}} \, e^{\frac{2 \pi i d}{\lambda} \left( 1 - \tilde{n} \right)},
\end{equation}
where $\overline{E}_{\mathrm{sam}}\left( \lambda \right)$ and $\overline{E}_{\mathrm{ref}}\left( \lambda \right)$ are respectively the discrete Fourier transforms of the first-pass and reference pulses shown in Fig.~\ref{FIG_THZ_TDS}a. The amplitude and phase of these signals are presented in the inset of Fig.~\ref{FIG_THZ_TDS}a. Within the frequency range of $0.3$--$1\, \mathrm{THz}$, the samples measured exhibit almost negligible loss. Instead of reporting the small $\kappa$ values that are subject to measurement noise, upper bounds for the extinction coefficient are listed in Table~\ref{TABLE_1} which are calculated assuming zero reflection of the THz beam off the sample, i.e.,
\begin{equation}\label{UPPER_BOUND_FOR_ATTENUATION_CONSTANT}
n = 1 \quad \Rightarrow \quad \kappa_{\mathrm{max}} = \frac{\lambda}{2 \pi d} \ln{\left| \frac{\overline{E}_{\mathrm{sam}}\left(\lambda\right)}{\overline{E}_{\mathrm{ref}}\left(\lambda\right)} \right|}.
\end{equation}
The refractive index and attenuation constant extracted via Eqs.~(\ref{TRANSMISSION_MODE_THZ_TDS})--(\ref{UPPER_BOUND_FOR_ATTENUATION_CONSTANT}) do not exhibit any considerable temperature-dependence within $5$--$295\,\mathrm{K}$. Moreover, the extracted results do not show any strong frequency-dependence within $0.3$--$1\, \mathrm{THz}$, and therefore, the spectrally-averaged results are presented in Table~\ref{TABLE_1}.
\begin{table}[H]
\centering
\caption{The refractive index, $n$, and the upper limit of attenuation constant, $\kappa_{\mathrm{max}}$, of annealed and untreated (S)GGG, respectively obtained using Eqs.~(\ref{TRANSMISSION_MODE_THZ_TDS}) and (\ref{UPPER_BOUND_FOR_ATTENUATION_CONSTANT}). The numbers are obtained via averaging the results over $0.3$--$1\, \mathrm{THz}$.}
\begin{tabular}{|l||c|c|}
	\hline
	Sample$\phantom{\frac{\frac{1}{2}}{\frac{1}{2}}}$		&	$\quad n \quad$	&	$\quad \kappa_{\mathrm{max}} \quad$		\\
	\hline
	Annealed GGG											&	$3.46$			&	$0.062$									\\
	\hline
	Annealed SGGG										&	$3.79$			&	$0.066$									\\
	\hline
	Untreated GGG										&	$3.49$			&	$0.059$									\\
	\hline
	Untreated SGGG										&	$3.80$			&	$0.066$									\\
	\hline
\end{tabular}
\label{TABLE_1}
\end{table}
The permittivity corresponding to the measured values of $n$ for all four substrates ranges from $12$ to $15$. This is consistent with the dielectric constant of $12.11$ measured along the $\langle 111 \rangle$ direction of crystalline GGG \cite{Lal_1977,10.1063/1.353856} and the measured polycrystalline dielectric constant of $11.9 \pm 1.9$ \cite{Young_Jin}. In contradistinction to the THz results, at a wavelength of $632.8\, \mathrm{nm}$ ($474\,\mathrm{THz}$), the refractive index of crsytalline GGG measured using the ellipsometry technique is reported to be $1.98 \pm 0.001$, independent of crystal orientation \cite{Belyaeva_2003}.

\subsection{Transmission measurement of biased substrates: Faraday rotation, Faraday ellipticity, and Verdet constant}\label{SUBSEC_II_III}

\begin{figure}[t!]
	\begin{center}
		\includegraphics[clip,width=0.98\columnwidth]{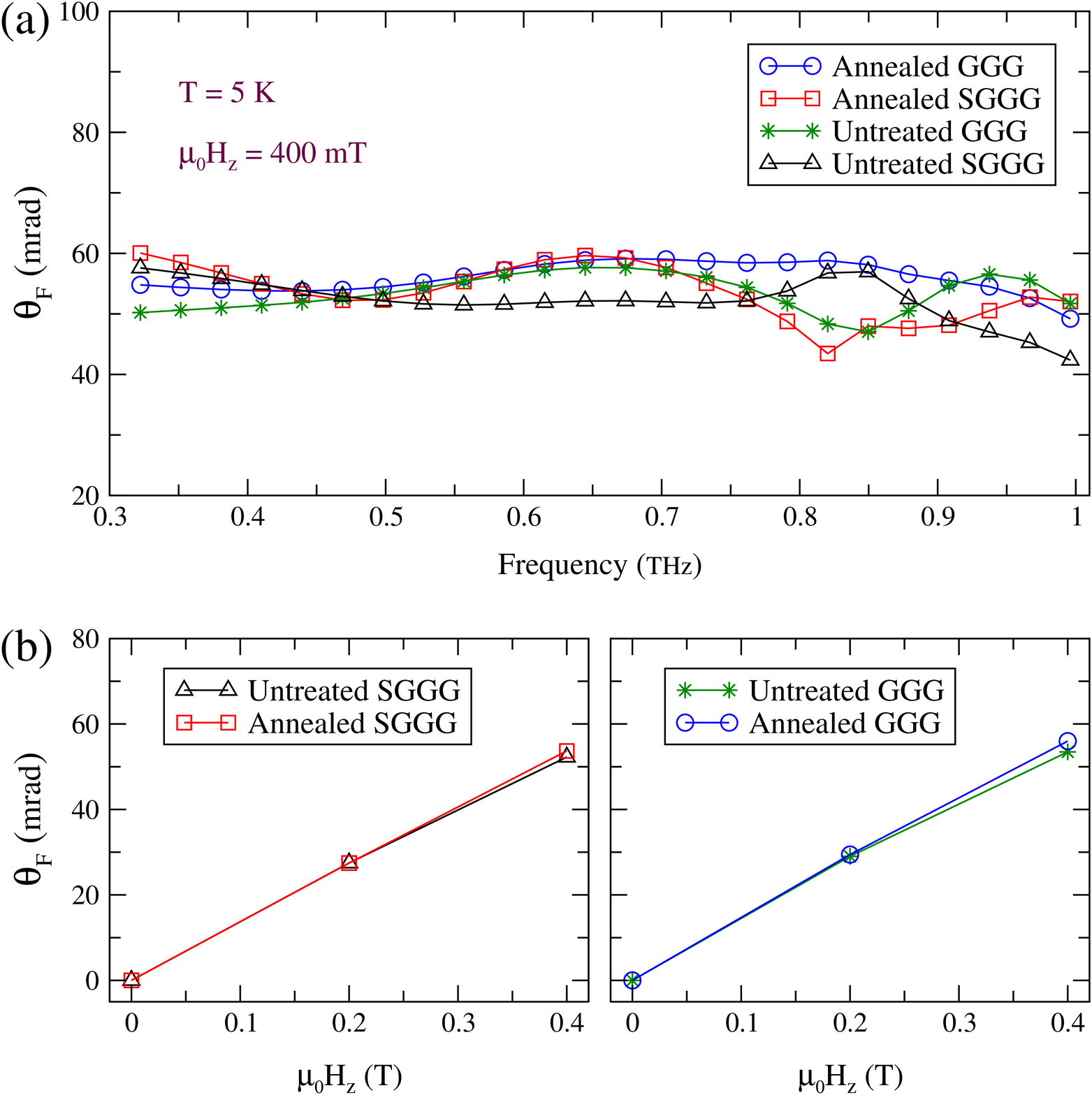}\\
		\caption{(a) Measured FR versus frequency for monocrystalline, annealed/untreated (S)GGG substrates at $T=5\,\mathrm{K}$ under an external bias of $B_{z}= 400\,\mathrm{mT}$. (b) and (c) The FR for annealed/untreated (S)GGG substrates obtained through spectral averaging over $0.3$--$1 \, \mathrm{THz}$.}	
		\label{FIG_SPECTRAL_FR}	
	\end{center}
\end{figure}

\begin{figure}[t!]
	\begin{center}
		\includegraphics[clip,width=0.98\columnwidth]{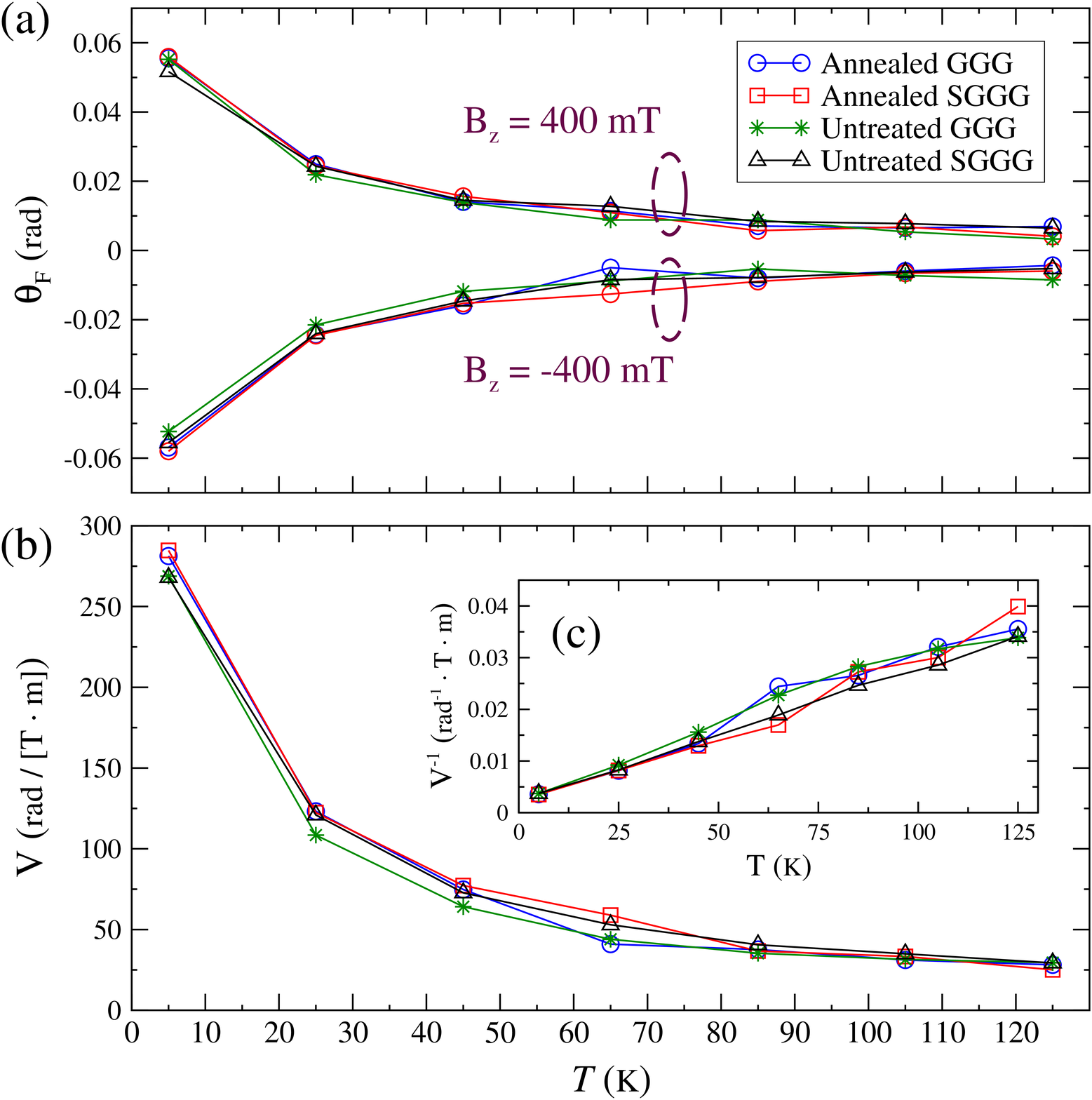}\\
		\caption{Measured FR versus temperature for annealed/untreated monocrystalline (S)GGG substrates when the magnetic bias is directed along $\langle 111 \rangle$ and $\langle \bar{1}\bar{1}\bar{1}\rangle$. The FR data are obtained via averaging their corresponding spectral FR data over $0.3$--$1\,\mathrm{THz}$ shown in Fig.~\ref{FIG_SPECTRAL_FR}a. The average of the real part of the two Verdet constants obtained from each of the cases of $B_{z} \!=\! \pm 400 \, \mathrm{mT}$ in panel (a). Inset (c) contains the inverse of the Verdet constant shown in panel (b).}
		\label{FIG_VERDET_CONSTANT_VERSUS_TEMPERATURE}
	\end{center}
\end{figure}

In Refs. \onlinecite{Pashkevich_2012,Galstyan_2015,Levy_2019}, wherein GGG is used as a substrate for substituted YIG films, the contribution of GGG to the overall FR of the composite system has been observed at $\lambda = 690\, \mathrm{nm}$, $530\,\mathrm{nm}$ and $532\, \mathrm{nm}$, respectively. Here, we present the FR measured within $0.3$--$1 \, \mathrm{THz}$ for a bare (S)GGG substrate. The detected signals shown in Figs.~\ref{FIG_THZ_TDS}b,~\ref{FIG_THZ_TDS}c are measured when the sample is biased with $\mu_{0} H_{z} = \pm 400\, \mathrm{mT}$ for wire-grid polarizer (WGP) angles of $\pm30^{\circ}$, respectively. Unlike the spectra shown in Fig.~\ref{FIG_THZ_TDS}a that are obtained in plain transmission measurement without magnet, the spectra in Figs.~\ref{FIG_THZ_TDS}b,c are from Faraday rotation setup with magnet. The presence of the magnet in the input and output path causes significant damping to the THz light intensity, and that is why the pulse profile in Figs.~\ref{FIG_THZ_TDS}b and \ref{FIG_THZ_TDS}c becomes much more broadened in comparison to the ones shown in Figs.~\ref{FIG_THZ_TDS}a. At a fixed external magnetic field, the polarization rotation of the transmitted THz pulse is obtained from the difference in signals detected at the two WGP angles
\begin{equation}\label{UPPER_BOUND_FOR_ATTENUATION_CONSTANT}
\theta_{\text{\tiny{F}}} = \arcsin{\left[\frac{E\left(30^{\circ}\right) - E\left(-30^{\circ}\right)}{2 \sin{\left(30^{\circ}\right)} \, E\left(0^{\circ}\right) }\right]},
\end{equation}
where $E\left(\pm 30^{\circ}\right)$ and $E\left(0^{\circ}\right)$ are the transmission field strength measured at WPG angles of $\pm 30$ and $0$ degrees, respectively. The pair of polarization rotation angles measured at magnetic fields with the same strength but opposite directions are compared and symmetrized to extract the components that are odd or even functions of the field. The odd component is attributed to the FR effect (Fig.~\ref{FIG_SPECTRAL_FR}), while the even component may originate from alternative field induced light modulation, such as the quadratic magneto-optical effect. Our measurement results are dominated by the FR related component (odd component), which is shown in Figure \ref{FIG_VERDET_CONSTANT_VERSUS_TEMPERATURE}a. As in Fig.~\ref{FIG_SPECTRAL_FR}, the FR data do not exhibit any strong dispersive behaviour within $0.3$--$1 \, \mathrm{THz}$, and therefore, we work with the spectrally-averaged FR values hereafter. No significant difference is observed comparing the annealed and untreated (S)GGG samples (Figs.~\ref{FIG_SPECTRAL_FR}b,c).

To report the FR and FE independent of bias and sample thickness, one can define a complex Verdet constant, $\tilde{V} \equiv \left[\theta_{\text{\tiny{F}}} + i\eta_{\text{\tiny{F}}}\right] / B_{z} d$, which can be obtained via normalizing the FR and FE by sample thickness, $d$, and external magnetic flux intensity, $B_{z}$. Since the upper limit of attenuation constant of all four samples is much smaller than its corresponding refractive index, the imaginary part of $\varepsilon_{g}$ is expected to be relatively small compared to its real counterpart. On the other hand, for a weak magnetic bias, $\varepsilon_{g}$ is expected to be smaller than its diagonal counterparts. In this way, the assumption of a negligible attenuation leads to the conclusion that the imaginary part of $\varepsilon_{g}$ should be negligible for weak amounts of magnetic bias. Implementing this assumption in Eq.~(\ref{IMAGINARY_PART_OF_THE_OFFDIAGONAL_COMPONENT_OF_THE_PERMITTIVITY_TENSOR}) yields $\eta_{\text{\tiny{F}}} \approx \frac{\kappa}{n} \theta_{\text{\tiny{F}}}$. This implies that the FE is expected to be much smaller than FR. As a result, the FE and attenuation constant are assumed negligible in our calculations. As a result, the Verdet constant is defined without the incorporation of FE, i.e.,
\begin{equation}\label{DEFINITION_OF_THE_VERDET_CONSTANT}
V \equiv \frac{\theta_{\text{\tiny{F}}}}{B_{z} d}.
\end{equation}
Since the magnetic fields ($< 400 \, \mathrm{mT}$) used in our experiments are much smaller than the typical saturation values ($\sim 10^{1} \,\mathrm{T}$) found in garnets \cite{Starobor_2011}, it is reasonable to treat the FR signal as a linear function of the field strength. This assumption is consistent with our field-dependent measurement results (Fig.\ref{FIG_SPECTRAL_FR}-b,c). Furthermore, the thermal expansion coefficient of GGG crystal, i.e., $\alpha_{T} = \frac{1}{a}\frac{\mathrm{d} a}{\mathrm{d} T}$, is of the order of $7 \times 10^{-7} \mathrm{K}^{-1}$ within the temperature range of $6$--$310\,\mathrm{K}$ \cite{Heinz_1972,osti_5740445}, with $a$ being the lattice constant. Therefore, the sample thickness is not expected to show a considerable temperature dependence, and the FR and Verdet constant are expected to exhibit the same temperature dependence.

As shown in Figure~\ref{FIG_VERDET_CONSTANT_VERSUS_TEMPERATURE}b, the measured Verdet constant is highly temperature dependent. Below $100 \, \mathrm{K}$, the Verdet constant within $0.3$--$1\, \mathrm{THz}$ significantly exceeds its reported values at visible-MIR (mid-infrared) frequencies, which range from $12.5$ to $22.3$ $\left(\mathrm{rad} \cdot \mathrm{T}^{-1} \cdot \mathrm{m}^{-1} \right)$ \cite{Starobor_2011,Novotny_2013}.

\section{Paramagnetic susceptibility and magneto-optic constant}\label{SEC:MOR}

\begin{figure}[t!]
	\begin{center}
		\includegraphics[clip,width=0.98\columnwidth]{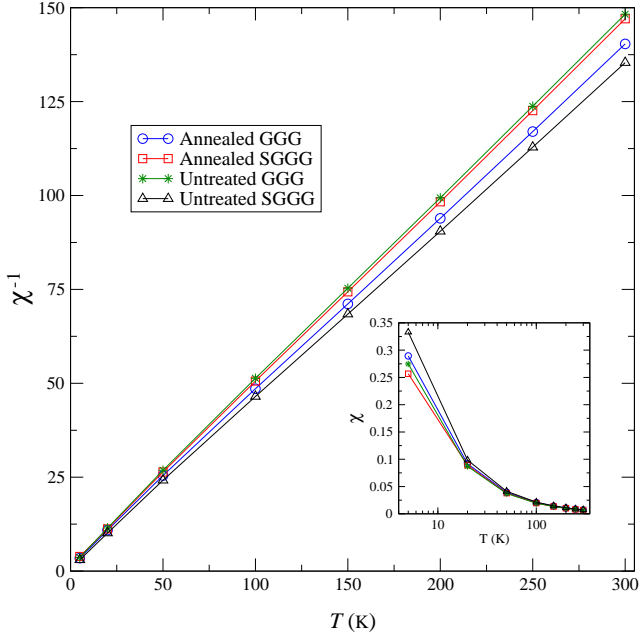}\\
		\caption{The measured DC magnetic susceptibility versus temperature for annealed/untreated monocrystalline (S)GGG substrates under a magnetic bias of $100\,\mathrm{mT}$ directed along the $\langle 111 \rangle$ direction. To distinguish the susceptibility data of these four samples from each other, a logarithmic scale has been used for the temperature axis in the inset panel.}
		\label{FIG_PARAMAGNETIC_SUSCEPTIBILTY}
	\end{center}
\end{figure}

Electrically-conductive and/or ferromagnetic material are known to exhibit gyrotropic response under magnetic bias \cite{Pozar}. However, (S)GGG does not fall in either of these categories; optical measurements of GGG indicate a bandgap of $5.66 \mathrm{eV}$ \cite{PODRAZA}. Therefore, the absence of an electron gas eliminates the possibility of a plasma-like gyrotropic response. On the other hand, our measurements indicate that the inverse of the Verdet constant increases linearly with temperature, as shown in Fig.~\ref{FIG_VERDET_CONSTANT_VERSUS_TEMPERATURE}c, with linear fitting parameters provided in Table~\ref{TABLE_2}.
\begin{table}[H]
\centering
\caption{Linear fit parameters for annealed/untreated monocrystalline (S)GGG substrates; the inverse of the real part of the Verdet constant is a linear function of temperature, i.e., $V = \frac{\beta}{T - T_{0}} \left[\frac{\mathrm{rad}}{\mathrm{T} \cdot \mathrm{m}}\right]$ (see Fig.~\ref{FIG_VERDET_CONSTANT_VERSUS_TEMPERATURE}c).}
\begin{tabular}{|l||c|c|c|}
	\hline
	Sample$\phantom{\frac{\frac{1}{2}}{\frac{1}{2}}}$ 	&	$\quad\beta  \left( \mathrm{K} \right)\quad $ 	&	$\;\; T_{0} \left( \mathrm{K} \right)\;\;$	\\
	\hline
	Annealed GGG										&	$36 \times 10^{2}$					&	$-8.2$										\\
	\hline
	Annealed SGGG									&	$34 \times 10^{2}$					&	$-1.4$										\\
	\hline
	Untreated GGG									&	$38 \times 10^{2}$					&	$-13$										\\
	\hline
	Untreated SGGG									&	$39 \times 10^{2}$					&	$-9.0$										\\
	\hline
\end{tabular}
\label{TABLE_2}
\end{table}
\noindent This behaviour is in agreement with the observation made in Refs. \onlinecite{Leycuras_1984, Liu_1995,PhysRevB.72.014415}. Moreover, the magnetic susceptibility of the (S)GGG substrates was measured using the vibrating-sample magnetometry (VSM) technique \cite{Foner_1959}: even at $T=5\,\mathrm{K}$, the sample magnetization did not show any hysteresis under sweeping the magnetic field $H_{z}$, and the positive slope of the $M$--$H$ linear curves indicated a paramagnetic response. This result is in agreement with the paramagnetic contribution of GGG to the perpendicular component of magnetization which has been observed as a linear background in the $M$--$H$ hystereses curves of composite XIG/GGG layered systems \cite{Pashkevich_2012,POPOVA_2013,Galstyan_2015,Sun_2015,Sokolov_2016,Stupakiewicz_2016,Avci_2016,Saito_2016,Quindeau_2017,Soumah_2018,Levy_2019}. As suggested by Fig.~\ref{FIG_PARAMAGNETIC_SUSCEPTIBILTY}, the temperature-dependence of the measured susceptibility data is described with the Curie-Weiss law
\begin{equation}\label{PARAMAGNETIC_SUSCEPTIBILTY}
\chi = \frac{C}{T - \Theta_{\mathrm{\scriptscriptstyle{CW}}}},
\end{equation}
where $ \Theta_{\mathrm{\scriptscriptstyle{CW}}}$ is the Curie-Weiss temperature and $C$ is the Curie-Weiss constant, which, by definition, is independent of temperature. The constants $ \Theta_{\mathrm{\scriptscriptstyle{CW}}}$ and $C$ have been obtained through linear fitting of the inversed susceptibility data and presented in Table~\ref{TABLE_3}.
\begin{table}[H]
\caption{Curie-Weiss parameters obtained through fitting the measured magnetic susceptibility for annealed/untreated monocrystalline (S)GGG substrates with the Curie-Weiss law given by Eq.~(\ref{PARAMAGNETIC_SUSCEPTIBILTY}). Comparison with the Curie-Weiss parameters obtained for a polycrystalline GGG sample in Ref. \onlinecite{PhysRevD.91.102004} shows reasonable agreement.}
\centering
\begin{tabular}{|l||c|c|c|}
	\hline
	Sample & $\phantom{\;\;}C \, \left(\mathrm{K}\right)\phantom{\;\;}$ & $\phantom{\;\;}\Theta_{\mathrm{\scriptscriptstyle{CW}}} \, \left(\mathrm{K}\right)\!\!\!\!\!\!\phantom{\frac{\frac{1}{2}}{\frac{1}{2}}}\;\;$\\
	\hline
	Annealed GGG								&	$2.2$	&	$-3.9$	\\
	\hline
	Annealed SGGG							&	$2.1$	&	$-3.7$	\\
	\hline
	Untreated GGG							&	$2.0$	&	$-3.9$	\\
	\hline
	Untreated SGGG							&	$2.2$	&	$-3.0$	\\
	\hline
	GGG (Ref. \onlinecite{PhysRevD.91.102004})	&	$2.0$	&	$-2.1$	\\
	\hline
\end{tabular}
\label{TABLE_3}
\end{table}
As shown in Fig.~\ref{FIG_PARAMAGNETIC_SUSCEPTIBILTY}, inverse susceptibility also increases linearly with temperature. However, $V / \chi$ is temperature dependent, which has been reported for the visible-frequency Verdet constant of the paramagnetic insulators such as NdF$_{3}$, PrF$_{3}$ and CeF$_{3}$ \cite{PhysRev.46.17,Leycuras_1984,PhysRevB.41.749,Liu_1995}.
	
\begin{figure}[t!]
	\begin{center}
		\includegraphics[clip,width=0.99\columnwidth]{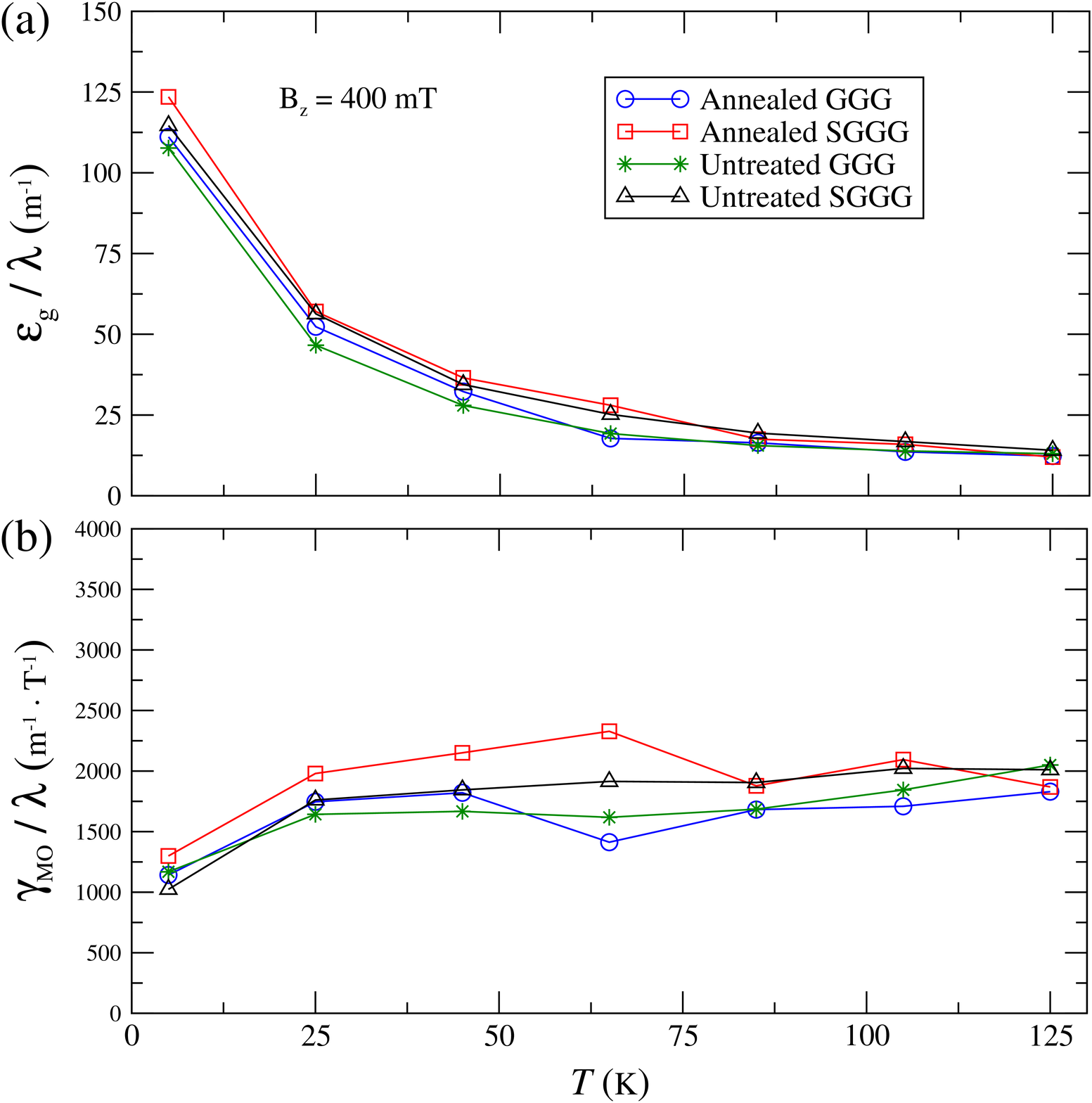}\\
		\caption{(a) The off-diagonal component of permittivity tensor of the annealed/untreated (S)GGG substrates obtained from Eq.~(\ref{REAL_PART_OF_THE_OFFDIAGONAL_COMPONENT_OF_THE_PERMITTIVITY_TENSOR}) using the measured refractive index, susceptibility and FR (for $\mu_{0} H_{z} = 400\, \mathrm{mT}$). (b) The magento-optical constant of the annealed/untreated (S)GGG substrates obtained using the measured refractive index, Verdet constant and susceptibility in Eq.~(\ref{THE_MAGNETO_OPTICAL_CONSTANT}). To remove the frequency dependence, the results are normalized by wavelength (in meters). Since the attenuation constant and FE are assumed to be zero within $0.3$--$1\,\mathrm{THz}$, both $\varepsilon_{g}$ and $\gamma_{\scriptscriptstyle{\mathrm{MO}}}$ are purely real.}
		\label{FIG_MAGNETO_OPTICAL_CONSTANT_VERSUS_TEMPERATURE}
	\end{center}
\end{figure}

The real part of $\varepsilon_{g}$ is computed for $\mu_{0} H_{z} = 400\,\mathrm{mT}$ via Eq.~(\ref{REAL_PART_OF_THE_OFFDIAGONAL_COMPONENT_OF_THE_PERMITTIVITY_TENSOR}) , and the results normalized by wavelength presented in Fig.~\ref{FIG_MAGNETO_OPTICAL_CONSTANT_VERSUS_TEMPERATURE}a. Since (S)GGG is an insulator, its magnetic response leads to the assumption of $\varepsilon_{g}$ being proportional to the DC magnetization: the magneto-optical (MO) response model \cite{Kirilyuk_2010,Tsai_2015,Rachid_2017,Subkhangulov}, i.e.,
\begin{equation}\label{MAGNETO_OPTICAL_RESPONSE}
\varepsilon_{g} = \mu_{0} \, \gamma_{\scriptscriptstyle{\mathrm{MO}}} \, M_{z},
\end{equation}
with $\gamma_{\scriptscriptstyle{\mathrm{MO}}}$ being the MO constant of the medium, $M_{z}$ denoting the projection of the DC magnetization vector onto the direction of propagation ($z$), and $\mu_{0}$ being the permeability of free space. In Ref. \onlinecite{Subkhangulov}, the assumption of MO response has been applied to TGG (Tb$_{5}$Ga$_{3}$O$_{12}$), which has a similar chemical and crystallographic structure to GGG.

The microscopic origin of the MO response is explained by the rotation of the excited dipolar currents as a result of the asymmetry of the electronic wave functions induced by the spin-orbit interaction \cite{Argyres_1995}. Combining Eqs.~(\ref{REAL_PART_OF_THE_OFFDIAGONAL_COMPONENT_OF_THE_PERMITTIVITY_TENSOR}), (\ref{IMAGINARY_PART_OF_THE_OFFDIAGONAL_COMPONENT_OF_THE_PERMITTIVITY_TENSOR}), (\ref{DEFINITION_OF_THE_VERDET_CONSTANT}) and (\ref{MAGNETO_OPTICAL_RESPONSE}), and assuming linear response between the magnetization and applied field, yields
\begin{equation}\label{THE_MAGNETO_OPTICAL_CONSTANT}
\gamma_{\scriptscriptstyle{\mathrm{MO}}} \cong \frac{\lambda}{\pi} \frac{\tilde{n}}{\sqrt{1 + \chi}} \frac{V}{\chi}.
\end{equation}

\noindent The $\sqrt{1 + \chi}$ factor in Eq.~(\ref{THE_MAGNETO_OPTICAL_CONSTANT}) can be traced back to the LHCP and RHCP refractive indices given by Eq.~(\ref{LHCP_RHCP_REFRACTIVE_INDICES}). Since $M_{z}$ in Eq.~(\ref{MAGNETO_OPTICAL_RESPONSE}) is purely real, $\gamma_{\scriptscriptstyle{\mathrm{MO}}}$ is required to be complex. However, as mentioned in Sec.~\ref{SUBSEC_II_III}, the imaginary part of $\varepsilon_{g}$, and therefore that of $\gamma_{\scriptscriptstyle{\mathrm{MO}}}$, is negligible within $0.3$--$1\,\mathrm{THz}$. The real part of $\gamma_{\scriptscriptstyle{\mathrm{MO}}}$, normalized by wavelength, is presented in Fig.~\ref{FIG_MAGNETO_OPTICAL_CONSTANT_VERSUS_TEMPERATURE}b, and appears to be nearly temperature-independent, except at low temperatures. The mechanism behind this is unclear to us.

It is worthwhile to mention that for magnetic biases as strong as $50 \, \mathrm{T}$, the Faraday rotation of (S)GGG substrates is expected to saturate with respect to magnetic bias due to the diamagnetic contribution of the oxygen-gallium bonds to the overall magnetization; a response which has been experimentally reported in Ref. \onlinecite{Levitin2002} for the case of TGG. As a result, the relation given by Eq.~(\ref{THE_MAGNETO_OPTICAL_CONSTANT}) may not be applicable to the case wherein the sample is subject to strong magntic bias.

\section{Summary and conclusions}\label{SEC:S_AND_C}

The gyroelectric permittivity tensor of annealed/untreated (S)GGG substrates is determined in the frequency range $0.3$--$1\,\mathrm{THz}$ and the temperature range $5$--$295\,\mathrm{K}$ using FR, magnetic susceptibility and refractive index measurements, whereas the ellipticity and absorption were found to be negligible. The Verdet and magneto-optic constants have been determined, and it was found that the diagonal elements do not exhibit any frequency dependence, and the off-diagonal elements are proportional to wavelength. The latter comment follows from Eqs.~(\ref{REAL_PART_OF_THE_OFFDIAGONAL_COMPONENT_OF_THE_PERMITTIVITY_TENSOR}--\ref{IMAGINARY_PART_OF_THE_OFFDIAGONAL_COMPONENT_OF_THE_PERMITTIVITY_TENSOR}), and the observation that both the refractive index and the Faraday rotation are frequency-insensitive within the considered frequency range.

Large Verdet constants approaching $300 \, \mathrm{rad}/[\mathrm{T}\cdot\mathrm{m}]$ are found in these paramagnetic materials at low temperatures. Such effect likely originates from the large magnetic permeability associated with the high-spin state of the Gd$^{3+}$ ions and the sizable magneto-optic constant. Future first-principle calculation and material modeling are called for to elucidate the detailed microscopic mechanism that gives rise to strong magneto-optic responses. The large Faraday rotation observed is insensitive to cation substitute and thermal treatment. The robustness of the strong magneto-optical effect, in conjunction with its broadband characteristics and the negligible loss found in the material, make (S)GGG wonderful candidates for making cryogenic THz isolators and circulators. As important substrate materials used for magnetic garnet thin film growth, the THz properties of (S)GGG systematically characterized in this work will also provide important information that are critical for the future development of garnet heterostructures based spintronic and magneto-optic devices.

\begin{acknowledgments}
Funding for this research was provided by the National Science Foundation under grant number EFMA-1741673. M. S. would like to thank Rostislav Mikhaylovskiy and Igor Ivanov for their insight on the MO response of TGG and the thermal expansion coefficient of GGG, respectively.  
\end{acknowledgments}

M. S. and F. S. contributed equally to this work.

\bibliography{REFERENCES}
\end{document}